\documentclass{article}

\usepackage{arxiv}

\usepackage[section]{placeins}
\usepackage{float}
\usepackage{url}

\usepackage[lofdepth,lotdepth]{subfig}

\usepackage[utf8]{inputenc} 
\usepackage[T1]{fontenc}    
\usepackage{hyperref}       
\hypersetup{colorlinks, citecolor=green, linkcolor=red, urlcolor=blue}
\usepackage{url}            
\usepackage{booktabs}       
\usepackage{amsfonts}       
\usepackage{nicefrac}       
\usepackage{microtype}      
\usepackage{lipsum}
\usepackage{graphicx}
\graphicspath{ {./images/} }
\usepackage{subfig}

\usepackage{array,multirow}

\usepackage{algorithm} 
\usepackage[noend]{algpseudocode} 

\usepackage{tikz}

\usepackage{hyperref}
\hypersetup{
    colorlinks=true,
    linkcolor=blue,
    filecolor=magenta,      
    urlcolor=cyan,
}

\usepackage{cite}

\usepackage{stackengine}

\usepackage{amsmath}
\usepackage{graphicx}

\def\NoNumber#1{{\def\alglinenumber##1{}\State #1}\addtocounter{ALG@line}{-1}}
\title{Simultaneous Quantum Machine Learning Training And Architecture Discovery}

\author{
 Dominic Pasquali \\
  Department of Physics\\
  University California Santa Cruz\\
  Santa Cruz, CA 95064 \\
  }
  
\begin{document}
\maketitle
\begin{abstract}
With the onset of gated quantum machine learning, the architecture for such a system is an open question. Many architectures are created either ad hoc or are directly analogous from known classical architectures. Presented here is a novel algorithm which learns a gated quantum machine learning architecture while simultaneously learning its parameters. This proof of concept and some of its variations are explored and discussed.
\end{abstract}

\section{Introduction}

In classical machine learning a machine learning practitioner must consider the model architecture and how to train that architecture. Model architecture and training is done either ad hoc or (more commonly) copied by looking at past architectures and training methods that were successful for similar problems, implementing those past architectures, and then making adjustments to the architecture and/or training method to account for differences between past problems and the problem at hand when fine tuning the trained model \cite{CurrentML}.

With quantum machine learning, all of the above problems are still present. Instead of classical parameterized layers there are quantum circuits with learned parameters; and instead of having known quantum circuit models for common problems, such models are created for a particular problem, dataset, or set of features. Therefore quantum machine learning practitioners, like their classical machine learning counterparts, are left to discover when to use which gates and how to efficiently train their quantum circuit. More recently, quantum machine learning architectures are analogous to or leverage techniques similar to classical architectures, including CNNs and transfer learning \cite{mari2019transfer, Cong_2019}. To summarize, the two main challenges that dominate quantum machine learning include model architecture creation (including gate selection) and model training (e.g. training gate parameters).

Presented in this paper is an algorithm which attempts to satisfy both challenges. The algorithm centers around the core principles that the product of unitary matrices can be expressed as a single unitary matrix, and the quantum logic gates Rx, Ry, Rz, and CNOT form a universal set of all quantum operators which can be performed on gated quantum systems\cite{Eleanor, vathsan, NielsenChuang2011}. The presented method begins with the assumption that there exists a unitary matrix which maps the desired input to the desired output. A matrix is learned and then acted on by an operation which transforms it into a unitary matrix. After a satisfactory criteria (e.g. accuracy, F1 score, or the like) is reached, a decomposition method takes in the unitary matrix and returns a sequence of gates with learned parameters whose net product reflects the behavior of the learned unitary matrix. Therefore at completion the algorithm returns the learned model architecture (i.e. the order which quantum gates act on which wires) and the learned parameters of those quantum gates (e.g. the rotation angles for rotation gates). This sequence of gates and parameters can be implemented on any quantum computer capable of implementing a universal set of quantum logic gates. Sequences of these gates can be recombined into well-known equivalent gates \cite{Eleanor, vathsan, NielsenChuang2011} thereby reducing the total number of operations that a quantum circuit will implement. 
Limitations of this algorithm will be discussed where appropriate.

\section{Basics \& Algorithm}

\subsection{Basics}
\label{sec:Basics}

Recall that all possible quantum gate operations on a collection of quantum wires can be represented as a series of tensor products which can be expressed as a single square unitary matrix $U$. Assume that each element in the matrix $U$ has both a real and complex part, resulting in each element of the matrix taking the form $x + iy$, where $x$ is the real part of the matrix element and $y$ is the imaginary part of the matrix element. Therefore the matrix $U$ can be expressed as $U = X+iY$, where $X$ and $Y$ are both real square matrices. It follows that the matrix $U$ will consist of: $2^{2n}\text{ real elements} + 2^{2n}\text{ imaginary elements} = 2(2^{2n}) = 2^{2n+1}$ total individual elements with $2^n \times 2^n = 4^n = 2^{2n}$ total matrix elements, where $n$ is the number of quantum wires.

Remember that there exist matrix decomposition methods, which when applied to a complex square matrix $M$ will yield a unitary matrix $U$ and a second matrix $R$ (the characteristics of the matrix $R$ depends on the chosen method of matrix decomposition)\cite{qrOne, qrTwo, schur1, schur2, schur3, polar1, polar2}. The set of these decomposition algorithms will be referred to as MU, where MU stands for matrix to unitary decomposition.

Also recall that algorithms exist that decompose a unitary matrix into elementary quantum gates\cite{PhysRevA.93.032318, Shende_2006, knill1995approximation, li2012decomposition, mottonen2005decompositions}. The set of these decomposition algorithms will be referred to as QG, where QG stands for quantum matrix to gate decomposition.


\subsection{The Algorithm}
\label{sec:Algorithm}
A mapping from an input to an output is desired. This input information is encoded into the quantum circuit. If the information is quantum in nature then this encoding is already complete, however if the information is classical then a classical to quantum encoding method may be chosen\cite{PL, TFQ, lloyd2020quantum}. 

A general unitary matrix $M$, where each element of $M$ is a learned parameter, acts as a whole on the states of the quantum circuit. For example in the matrix in \autoref{eq:M}, the variables $a$, $b$, $c$, $d$, $e$, $f$, $g$, and $h$ (or alternatively $j$, $k$, $l$, $m$, $n$, $p$, $q$, and $r$) are learned. Note that some or all of the circuit's initialized states are passed to $M$, as chosen by the user.

\begin{equation}
M = \left[ \begin{array}{ccc} a+ib & ... & c+id  \\ ... & ... & ... \\ e+if & ... & g+ih \end{array}\right] = \left[ \begin{array}{ccc} je^{ik} & ... & le^{im}  \\ ... & ... & ... \\ ne^{ip} & ... & qe^{ir} \end{array}\right]
\label{eq:M}
\end{equation}


Measurements on the wires of interest are taken. These measurements constitute the output of the quantum circuit. The loss function is calculated, comparing the output of the network and the target value. Backpropagation then calculates how much the parameters in the circuit should change; these parameters are then updated in matrix $M$. After updating the values in $M$, MU decomposition is then enacted on $M$ which then yields matrix $U$. The square complex unitary matrix $U$ then replaces the matrix $M$.

The above steps are repeated, except now the elements of $U$ are learned, backpropagated, and updated.

After the data is sufficiently iterated through (or until desired performance metrics have reached a satisfactory level), a learned square complex unitary matrix, $V$, remains. QG decomposition is performed on matrix $V$ such that a series of elementary quantum gates and their parameters needed to represent the actions of the trained matrix $V$ are extracted. (Recall that the product of the gates constitutes the needed quantum architecture for the quantum model architecture in terms of a universal set of quantum logic gates.) The resulting series of gates also contains with them their parameterized values. Therefore the returned series of gates is the quantum circuit architecture, and the returned parameters of those gates forms the fully trained quantum model.

\subsection{Algorithm Variations}
\label{sec:AlgorithmVariations}
There are many different variations of this algorithm, some of which could have varying effects on the result of the output. Some of these variations are mentioned below.

While executing the algorithm, the learned matrix can be checked after each update step to see if it is unitary. If the transformed matrix is unitary, then it would be preferable not to perform MU decomposition on the learned matrix since preserving as much of the update step information as possible would aid in learning the correct matrix.

In Sections \ref{sec:Basics} and \ref{sec:AlgorithmVariations} MU decomposition represented a set of decomposition methods to transform a non-unitary matrix to a unitary matrix, all of which includes a single complex square unitary matrix (e.g. QR, Schur, and Polar decomposition)\cite{qrOne, qrTwo, schur1, schur2, schur3, polar1, polar2}. Such a decomposition method is chosen by the user.

QG decomposition represents a set of decomposition methods which extract what gates could represent the final learned unitary matrix $V$. These methods all result with a decomposition that includes Quantum Shannon Decomposition (QSD) \cite{PhysRevA.93.032318, Shende_2006}, Knill Decomposition \cite{knill1995approximation}, and others \cite{li2012decomposition, mottonen2005decompositions}. The decomposition method to extract the gate representation of the learned unitary matrix is also chosen by the user.

In the model in Section \ref{sec:Algorithm}, the desired input information went straight into a quantum circuit. It is also possible to construct a hybrid model such that a classical neural network is attached on either side or both sides of the quantum circuit as to simultaneously train and learn the matrix $V$ and the weights of the classical neural network. In some cases a series of different quantum, classical, quantum and then classical, or classical and then quantum layers may be desired.

In assembling the quantum circuit to implement the algorithm, there may be reason for the placement of ad hoc quantum gates in the circuit for quantum data processing. A practitioner may also decide to apply and learn more than one general unitary matrix for their quantum circuit, and such unitary matrices may or may not overlap with wires on which they act.


Ancilla bits may be added to the quantum circuit being used. The proposed learned matrix approach allows the network to learn the interactions and operations between particular values as well as the interactions between ancilla bits (e.g. $|0>$, $|1>$, or otherwise). Such interactions become important especially if there exists an input of $m$ pieces of information but $n$ wires are measured, where $m > n$. Such a variation is of concern to methodologies that mimic QCNNs \cite{Cong_2019}.



\subsection{Other Hyperparameters, Constraints, And Limitations}

\begin{figure} 
    \centering
    \includegraphics[width=7.2cm]{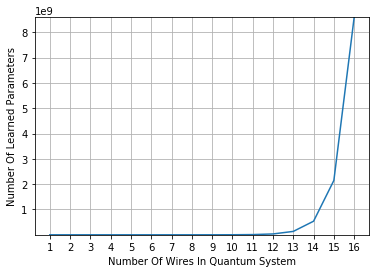}
    \caption{Number of qubits vs number of learned parameters.} 
    \label{fig:params} 
\end{figure}

While training or preparing for training, it helps to recall the total number of trained parameters. The number of learned elements in the proposed setup rises exponentially. If sixteen quantum wires are used then the model will learn $2^{(2 \times 16)+1} = ~8.6$ billion parameters, which quickly approaches the size of the largest state-of-the-art machine learning models as seen in \autoref{fig:params}. Therefore while using this methodology for current NISQ devices, practitioners need to be sensitive to the number of learned and stored parameters on classical computers.

\begin{figure} 
    \centering
    \includegraphics[width=14cm]{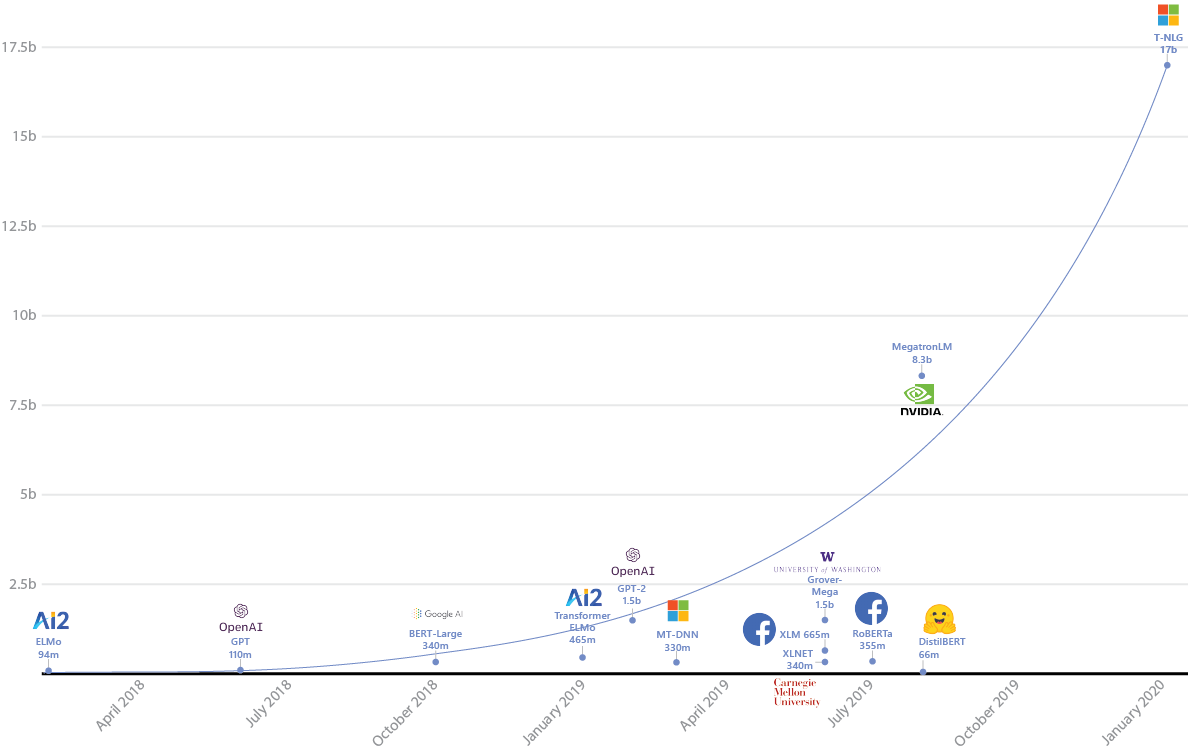}
    \caption{Large models in modern day machine learning research \cite{micro}.} 
    \label{fig:micro} 
\end{figure}



The proposed algorithm also has the drawback of needing to deconstruct the learned unitary matrix into individual quantum gates. The number of quantum gates needed to represent a unitary matrix scales in correlation with the type of applied deconstruction \cite{iten2019introduction}. If the accumulated noise from the needed gates for the unitary matrix representation reaches a level beyond acceptable tolerance, then the quantum machine learning practitioner may choose a different unitary matrix to quantum gate deconstruction algorithm (resulting in fewer deconstructed quantum gates) or choose an entirely different implementation. 

Another challenge is taking the gradient of the quantum circuit. Methods of taking the quantum gradient is an active area of research; recent developments propose a quantum gradient which can be solved for analytically \cite{crooks2019gradients,schuld2018evaluating}. In this paper the gradient of a quantum circuit is calculated via numerical differentiation (using the central difference) as defined in \autoref{eq:grad}, where $f$ is an expectation value dependent on $\theta$, which are gate parameters \cite{MultiParamGates, guerreschi2017practical}. Note that $\Delta \theta$ in \autoref{eq:grad} is a hyperparameter.

\begin{equation}
\nabla f(\theta) = \frac{d f(\theta)}{d \theta} \approx \frac{f(\theta + \frac{\Delta \theta}{2})- f(\theta - \frac{\Delta \theta}{2})}{\Delta \theta}
\label{eq:grad}
\end{equation}

\section{Experiment}

Despite the plethora of possible variations of this algorithm, presented here are three such variations.

\subsection{Software, Dataset, Data Preparation}
The two moons dataset from ScikitLearn \cite{sklearn_api} was chosen as a classification example. Pennylane \cite{bergholm2018pennylane} was used to simulate the quantum circuit. PyTorch \cite{NEURIPS2019_9015} supplied the backend machine learning and computational mechanics. 3200 training samples were used along with 400 testing samples and 400 validation samples, all of which had a noise level of 5\% with a batch size of 100 for each epoch. The ADAM optimizer was used with an initialized learning rate of 0.01. When implementing \autoref{eq:grad} to solve for the gradient, $\Delta \theta=\frac{\pi}{10}$ was used for both the real and imaginary parts of the matrix. To deconstruct the quantum unitary matrix, the UniversalQCompiler \cite{iten2019introduction} was used to implement QSD.

Since PyTorch does not support complex values, the proposed square complex matrix was separated into its real and imaginary parts. Such parts were then learned and recombined as necessary to form the square complex unitary matrix. It is for this reason that the gradient was calculated for both the real and imaginary matrices separately as opposed to calculating a single gradient for a single square complex matrix.

In the utilized architectures and their corresponding implementations, a set seed was used for each experiment to control for repeatable results.

\subsection{Model A}
\label{sec:full_U}

This is a hybrid classical-quantum model where a block $16\times16$ unitary matrix is simultaneously learned for all four wires.

\subsubsection{Architecture}

The following pipeline was chosen to demonstrate the proposed algorithm.

\floatname{algorithm}{Model A}
\def\NoNumber#1{{\def\alglinenumber##1{}\State #1}\addtocounter{ALG@line}{-1}}
\begin{algorithm}
\renewcommand{\thealgorithm}{}
\caption{Pipeline}\label{euclid}
\begin{algorithmic}[1]
\For{\text{ each data point in each epoch}}
\State The x and y coordinates of the single point is taken and input into the next layer
\State Classical linear dense layer with 2 inputs and 4 outputs
\State A hyperbolic tangent activation function acts on the 4 outputs
\State The output is then multiplied by $\frac{\pi}{2}$ to create new scaled outputs\footnotemark
\State The outputs are fed into a quantum circuit with 4 wires 
\State The quantum circuit takes each input and passes them into an Ry gate (as seen in \autoref{eq:RY}) on each wire
\State All of the wires are then acted on by a square complex unitary matrix
\State Each wire's expectation value of the Pauli-Z observable is taken
\State These measurements are fed into a dense layer with 4 classical inputs to 2 classical outputs
\State A hyperbolic tangent activation function acts on the 2 outputs
\State The output of each classical node is then compared against the truth value via the Cross Entropy loss function
\State After calculating the loss function, the parameters of the model are updated
\State The updated parameters now form a complex square matrix and this matrix undergoes Schur decomposition 
\State The resulting unitary matrix is kept and the other resulting matrix is then discarded
\If{\text{ satisfactory accuracy is attained}} \textbf{break}
\EndIf
\EndFor
\State QSD then operates on the last generated square complex unitary matrix to yield a sequence of quantum gates with their corresponding wire number(s) (which wire(s) they act on) and their rotation parameter where applicable (e.g. CNOT gates would not have rotation parameters since they only act on given wires while rotation gates will have a wire value parameter and a rotation value parameter)
\end{algorithmic}
\end{algorithm}

\footnotetext{Scaling the inputs into the quantum circuit is a crucial operation. Recall that values on the Bloch sphere can only rotate between $0$ and $2\pi$. If the values (in radians) inserted into the quantum circuit ignore this cyclic property, then the algorithm may run afoul with respect to overlapping projected values. For example, say that two different values are plugged into the quantum circuit via encoding through an Ry gate rotation. Let one value be 1.35 which belongs to Datum 1 and the other value be 13.91637 which belongs to Datum 2. Since these values are both are in radians they are mapped onto the Bloch sphere at the exact same place which leads to erroneous training of the model. Therefore imposing a hyperbolic tangent function on the values ensures that all of the inputs to the quantum circuit are between $1$ and $-1$ and by multiplying all inputs by $\frac{\pi}{2}$ allows the inputs of the quantum circuit to lie between $\frac{\pi}{2}$ and -$\frac{\pi}{2}$ before being given an Ry rotation on the Bloch sphere. If the data was more sparse, there would be incentive to project the data between a larger range of values. Such a scaling must be chosen by the user or learned. Therefore it is crucial for a practitioner to ensure that all data and information encoded into a quantum circuit remain unique. See Section \ref{sec:learned_full_U} for an attempt at learning this scaling parameter.}

\begin{equation}
R_y(\phi) = \left[ \begin{array}{cc} \cos{\frac{\phi}{2}} & -\sin{\frac{\phi}{2}}  \\ \sin{\frac{\phi}{2}} & \cos{\frac{\phi}{2}} \end{array}\right]
\label{eq:RY} 
\end{equation}

\subsection{Model B}
\label{sec:4_Q_wires_8_terms}

This model is nearly identical to Model A in Section \ref{sec:full_U}, with the exception that an individual complex unitary matrix ($2 \times 2$) is learned for each individual wire.

\subsection{Model C}
\label{sec:learned_full_U}

Note that in Model A in Section \ref{sec:full_U} the model projected all of the input data to be between $\frac{\pi}{2}$ to $-\frac{\pi}{2}$. This model is similar to Model A in Section \ref{sec:full_U} except it attempts to learn the scaling parameter, $\alpha$, in \autoref{eq:scaling}. Therefore the architecture of this model is the same as that in Section \ref{sec:full_U} with the exception that lines 4 and 5 in Model A's pipeline become "these four outputs are then multiplied by $\alpha\tanh{(x)}$, where $\alpha$ is a learned parameter that creates new scaled outputs."

\begin{equation}
\phi(x) = \alpha\tanh{(x)}
\label{eq:scaling} 
\end{equation}

\section{Results and Analysis}



\begin{figure}[h]
\centering
    \includegraphics[width=0.3\textwidth]{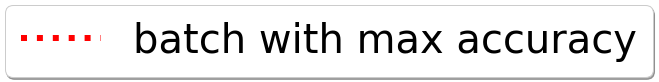}
    \label{fig:blank}

\hspace{2.0cm}
\stackunder[0pt]{
\includegraphics[width=0.1\textwidth]{ann/ModelA.png}}{}
\hspace{3.8cm} 
\stackunder[0pt]{
\includegraphics[width=0.1\textwidth]{ann/ModelB.png}}{}
\hspace{3.8cm} 
\stackunder[0pt]{
\includegraphics[width=0.1\textwidth]{ann/ModelC.png}}{}
\hspace{0.8cm}

\centering
\subfloat[text][Max accuracy = 98.8125\%]{
\includegraphics[trim=0 0 180 0,clip,width=0.335\textwidth]{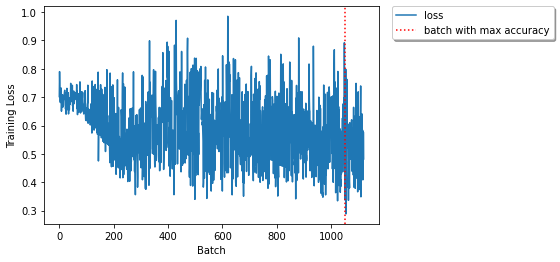}
\label{fig:full_U_seed0_trend_1}}
\subfloat[text][Max accuracy = 89.8125\%]{
\includegraphics[trim=0 0 180 0,clip,width=0.335\textwidth]{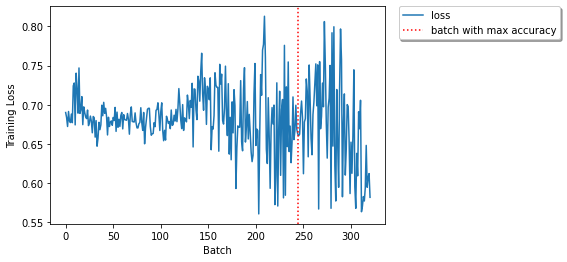}
\label{fig:full_U_seed0_indi_trend_1}}
\subfloat[text][Max accuracy = 89.375\%]{
\includegraphics[trim=0 0 173 0,clip,width=0.335\textwidth]{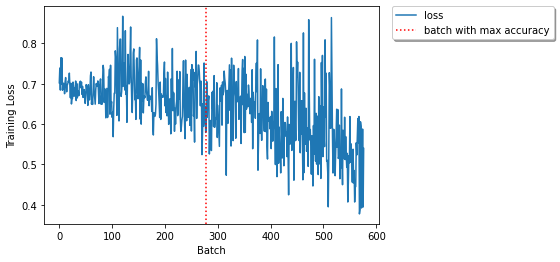}
\label{fig:full_U_seed0_learn_trend_1}}

\subfloat[text][Max accuracy = 98.8125\%]{
\includegraphics[trim=0 0 180 0,clip,width=0.335\textwidth]{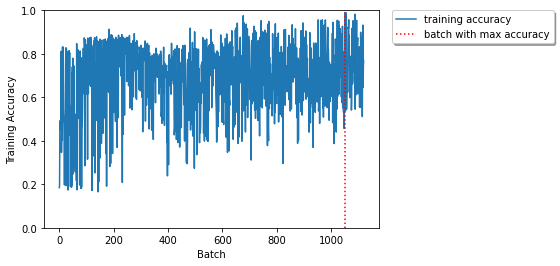}
\label{fig:full_U_seed0_trend_2}}
\subfloat[text][Max accuracy = 89.8125\%]{
\includegraphics[trim=0 0 180 0,clip,width=0.335\textwidth]{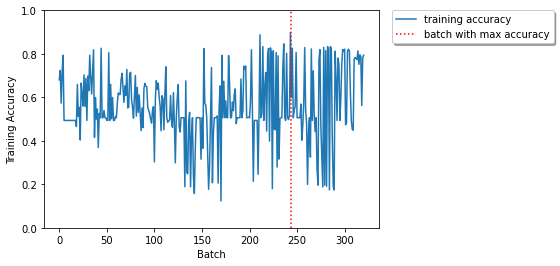}
\label{fig:full_U_seed0_indi_trend_2}}
\subfloat[text][Max accuracy = 89.375\%]{
\includegraphics[trim=0 0 173 0,clip,width=0.335\textwidth]{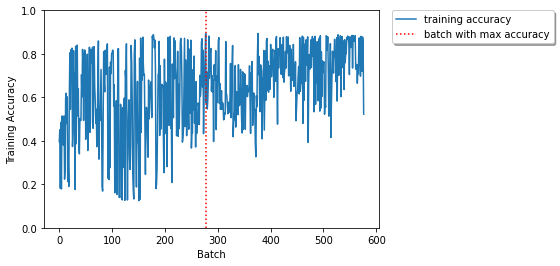}
\label{fig:full_U_seed0_learn_trend_2}}

\subfloat[text][Max accuracy = 98.75\%]{
\includegraphics[trim=0 0 180 0,clip,width=0.335\textwidth]{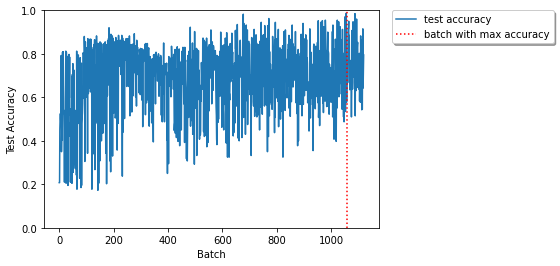}
\label{fig:full_U_seed0_trend_3}}
\subfloat[text][Max accuracy = 90\%]{
\includegraphics[trim=0 0 180 0,clip,width=0.335\textwidth]{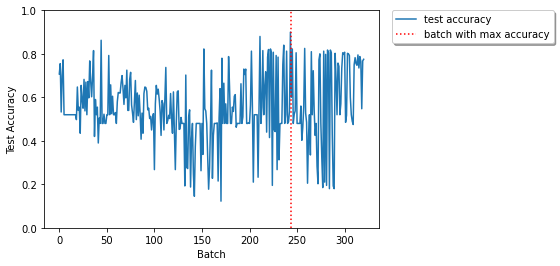}
\label{fig:full_U_seed0_indi_trend_3}}
\subfloat[text][Max accuracy = 91.5\%]{
\includegraphics[trim=0 0 173 0,clip,width=0.335\textwidth]{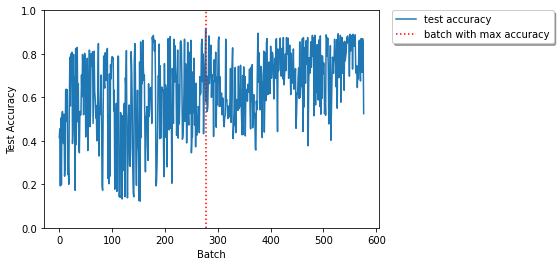}
\label{fig:full_U_seed0_learn_trend_3}}

\subfloat[text][Max accuracy = 99.25\%]{
\includegraphics[trim=0 0 180 0,clip,width=0.335\textwidth]{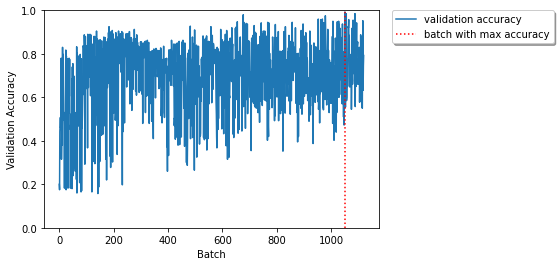}
\label{fig:full_U_seed0_trend_4}}
\subfloat[text][Max accuracy = 91.25\%]{
\includegraphics[trim=0 0 180 0,clip,width=0.335\textwidth]{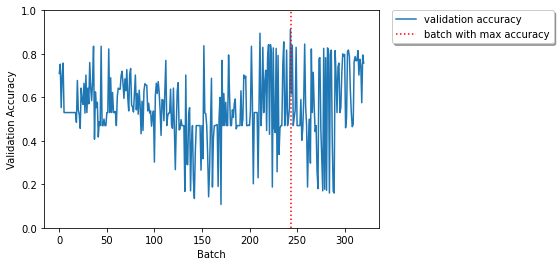}
\label{fig:full_U_seed0_indi_trend_4}}
\subfloat[text][Max accuracy = 91\%]{
\includegraphics[trim=0 0 173 0,clip,width=0.335\textwidth]{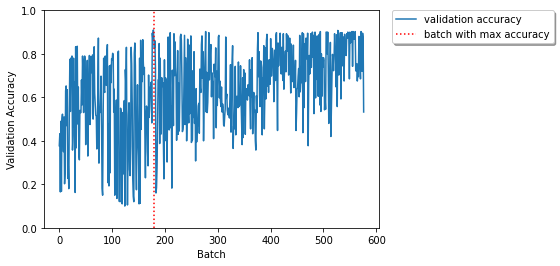}
\label{fig:full_U_seed0_learn_trend_4}}

\caption{Plots depicting loss and accuracy. The red line in the loss plot corresponds to the maximum training accuracy batch. Model A's training and validation plots resulted after 32 epochs + 27 batches. Model A's test plot resulted after 33 epochs. All of Model B's plots resulted after 7 epochs + 19 batches. Model C's training and test plots resulted after 8 epochs + 21 batches. Model C's validation plot resulted after 5 epochs + 18 batches.
}
\label{fig:full_seed0_trend}
\end{figure}

While all three models learn non-linear aspects of the two moons dataset, Model A reaches a better accuracy than Model B or C, as seen in \autoref{fig:full_seed0_trend}. Model A does reach the better accuracy while also needing more training batches to achieve such accuracy compared to the other two models. Model A and Model C appear to slowly rise in accuracy at about the same rate while Model B appears to reach its best accuracy almost stochastically. 

\begin{figure}[h]
\centering

\stackunder[0pt]{
\includegraphics[width=0.03\textwidth]{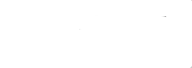}}{}
\stackunder[0pt]{
\includegraphics[width=0.1\textwidth]{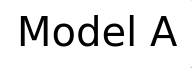}}{}
\quad\qquad\qquad\qquad\qquad\quad 
\stackunder[0pt]{
\includegraphics[width=0.1\textwidth]{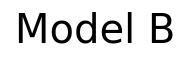}}{}
\quad\qquad\qquad\qquad\qquad\quad 
\stackunder[0pt]{
\includegraphics[width=0.1\textwidth]{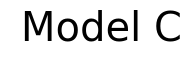}}{}

\stackunder[0pt]{
\raisebox{.4\height}{\includegraphics[width=0.03\textwidth]{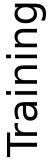}}}
\enskip
\subfloat[text][Total accuracy = 98.8125\%]{
\includegraphics[width=0.29\textwidth]{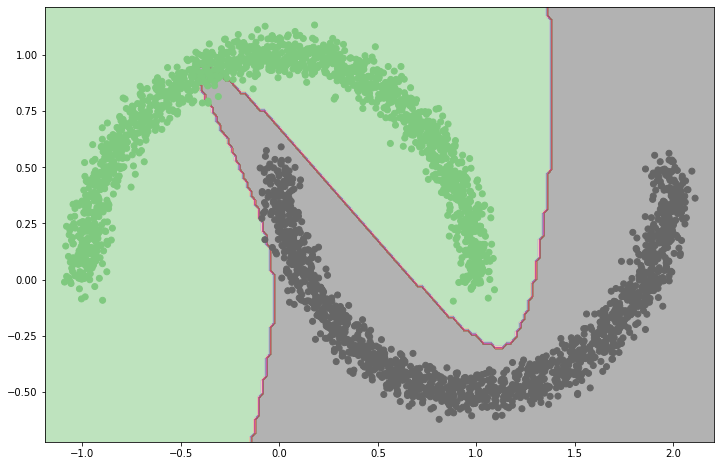}
\label{fig:full_U_seed0_1}}
\quad
\subfloat[text][Total accuracy = 89.8125\%]{
\includegraphics[width=0.29\textwidth]{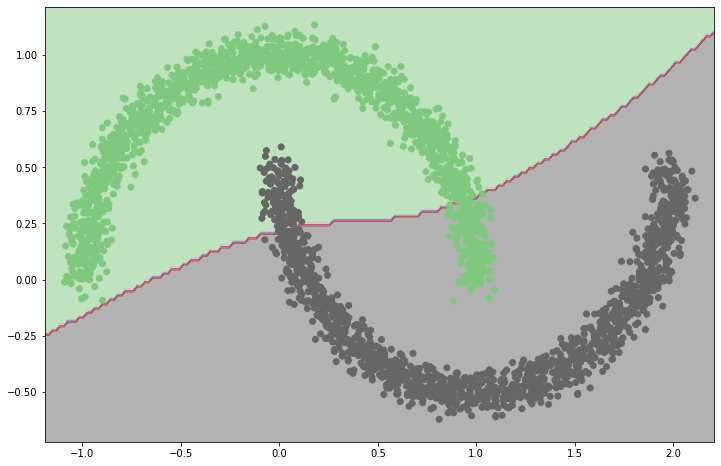}
\label{fig:4_wires_seed0_1}}
\quad
\subfloat[text][Total accuracy = 89.375\%]{
\includegraphics[width=0.29\textwidth]{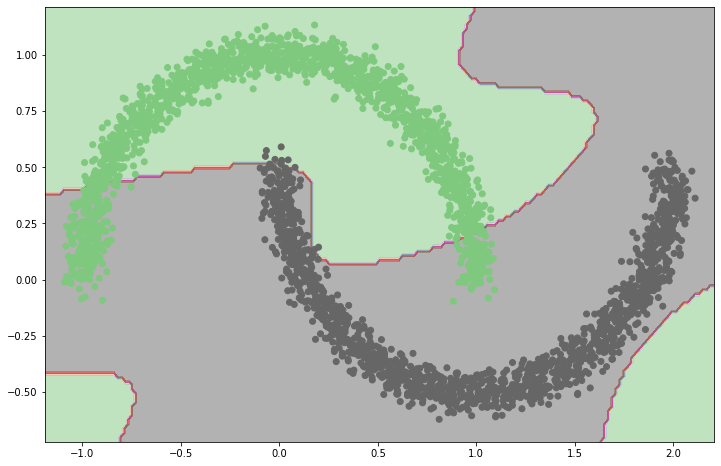}
\label{fig:full_U_seed0_learn_1}}

\stackunder[0pt]{
\raisebox{1.40\height}{\includegraphics[width=0.035\textwidth]{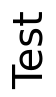}}}
\enskip
\subfloat[text][Total accuracy = 98.75\%]{
\includegraphics[width=0.29\textwidth]{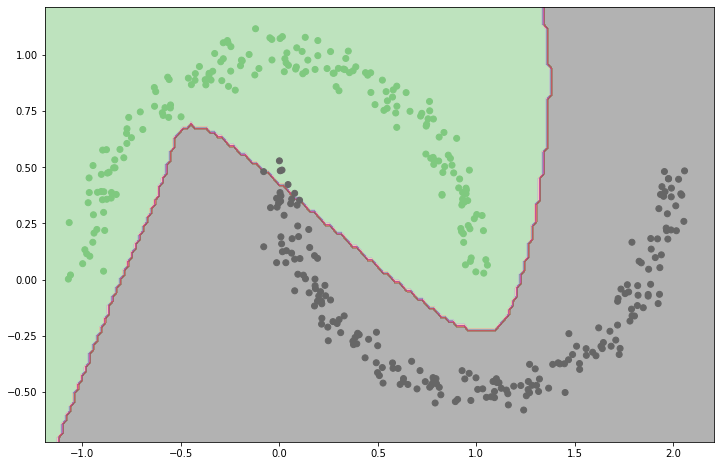}
\label{fig:full_U_seed0_4}}
\quad
\subfloat[text][Total accuracy = 90\%]{
\includegraphics[width=0.29\textwidth]{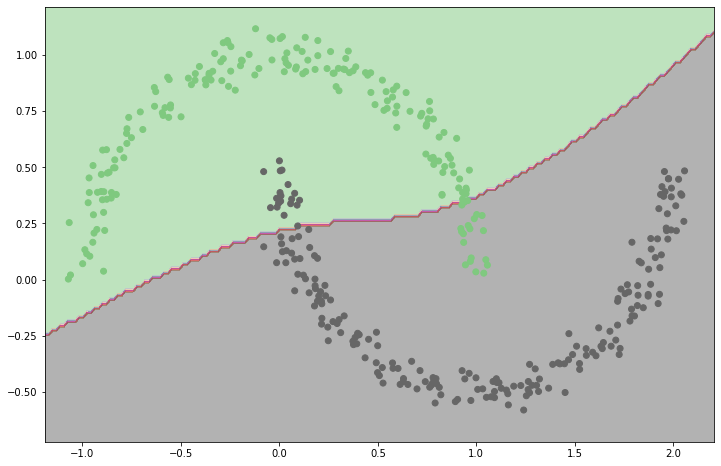}
\label{fig:4_wires_seed0_2}}
\quad
\subfloat[text][Total accuracy = 91.5\%]{
\includegraphics[width=0.29\textwidth]{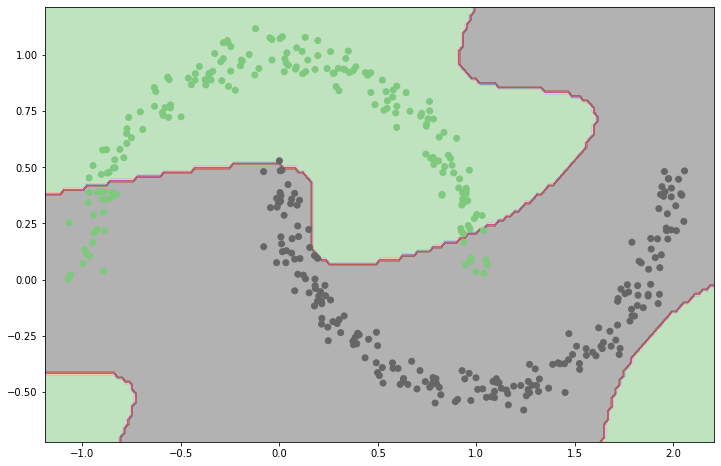}
\label{fig:full_U_seed0_learn_4}}

\stackunder[0pt]{
\raisebox{0.3\height}{\includegraphics[width=0.033\textwidth]{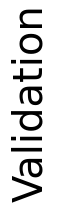}}}
\enskip
\subfloat[text][Total accuracy = 99.25\%]{
\includegraphics[width=0.29\textwidth]{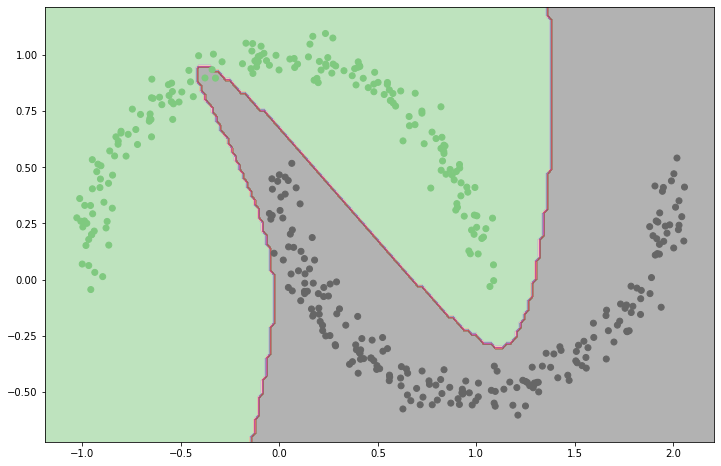}
\label{fig:full_U_seed0_7}}
\quad
\subfloat[text][Total accuracy = 91.25\%]{
\includegraphics[width=0.29\textwidth]{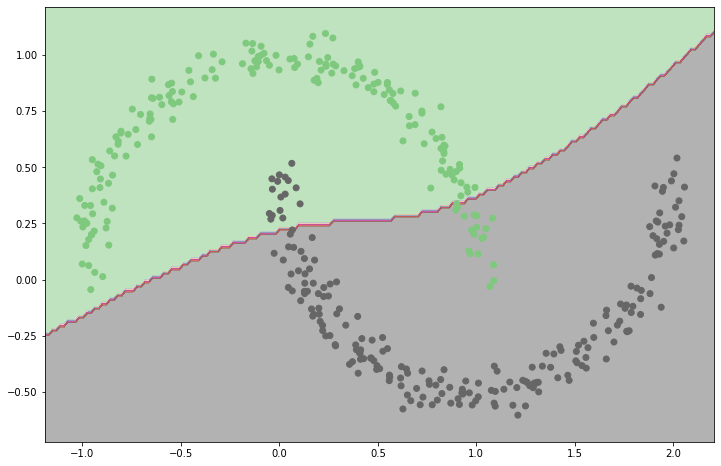}
\label{fig:4_wires_seed0_3}}
\quad
\subfloat[text][Total accuracy = 91\%]{
\includegraphics[width=0.29\textwidth]{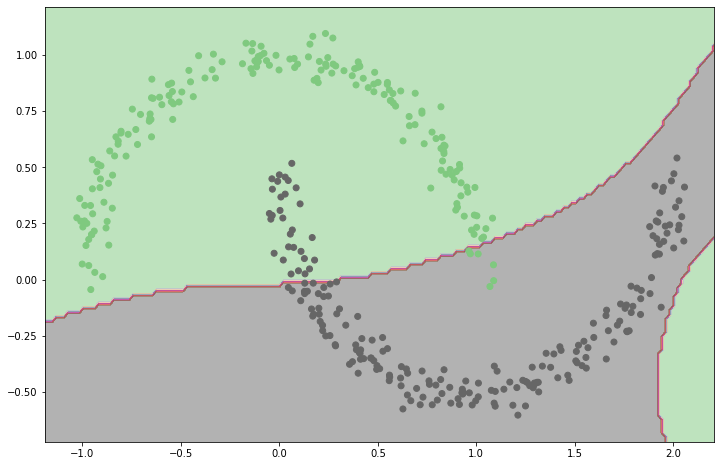}
\label{fig:full_U_seed0_learn_7}}

\caption{Plots of the contour decision boundary lines based on the learned unitary matrix. Model A's training and validation plots resulted after 32 epochs + 27 batches. Model A's test plot resulted after 33 epochs. All of Model B's plots resulted after 7 epochs + 19 batches. Model C's training and test plots resulted after 8 epochs + 21 batches. Model C's validation plot resulted after 5 epochs + 18 batches. 
}
\label{fig:full_U_seed0}
\end{figure}

Model B's learning (as seen in Figures\autoref{fig:full_U_seed0_indi_trend_1},\autoref{fig:full_U_seed0_indi_trend_2},\autoref{fig:full_U_seed0_indi_trend_3}, and\autoref{fig:full_U_seed0_indi_trend_4}) can be interpreted as the optimization for the input and the output per wire; however such optimization doesn't take into consideration the interaction between wires. Since a general unitary matrix for each specific wire is a single unitary matrix, this matrix can be expressed by three rotations around two axes \cite{NielsenChuang2011}. By independently learning the unitary matrix for each wire the algorithm learns the optimized rotations around the optimized axes. However when the accuracy for the learned matrices for each individual wire is compared against Model A in which a fully learned unitary is applied to all the wires simultaneously (as seen in Figures\autoref{fig:full_U_seed0_1},\autoref{fig:full_U_seed0_4}, and \autoref{fig:full_U_seed0_7}), the fully learned unitary matrix outperforms the individually learned matrices for each wire. This difference reflected in the corresponding plots emphasizes the importance to learn the full unitary matrix as applied to the entire circuit all at once, which allows the architecture to learn both the individual coupling terms and interaction terms on and between wires. This may be why the loss was less dramatic but the accuracy, on average, never increased for Model B. There is confidence that learning the coupling between wires is essential for drawing the optimal decision boundary and reaching a better accuracy.

Model A also efficiently maximizes of the number of learned parameters on the number of quantum wires available since every real and imaginary term in the matrix is learned independently.

In training Model C (as seen in Figures\autoref{fig:full_U_seed0_learn_trend_1},\autoref{fig:full_U_seed0_learn_trend_2},\autoref{fig:full_U_seed0_learn_trend_3}, and\autoref{fig:full_U_seed0_learn_trend_4}), learning the parameter $\alpha$ does not improve the overall accuracy despite having the exact same initialized hyperparameters as Model A. A different architecture or differently chosen hyperparameters may be needed to properly accommodate or tune this scaling. Which architectures and hyperparameters to use and their theoretical motivations to optimize Model C are beyond the scope of this paper.

When looking across the training, test, and validation plots for Models A, B, and C, (\autoref{fig:full_seed0_trend}) it is found that when the ideal accuracy has been reached, the network appears to jump out of the local minimum. This could happen for a number of reasons, for example the optimization steps used in the optimizer ADAM were so big that the network jumped out of the current (and best yet) local minima. A learning rate scheduler might have resolved this problem. This jumping out of the local minima might also be resolved with the utilization of a different algorithm to find the quantum gradient, since finite differences gives a coarse gradient which is dependent on the defined hyperparameter $\Delta \theta$. 

\begin{figure}[h]
\centering

\stackunder[0pt]{
\includegraphics[width=0.03\textwidth]{ann/blank2.png}}{}
\stackunder[0pt]{
\includegraphics[width=0.1\textwidth]{ann/ModelA.png}}{}
\quad\qquad\qquad\qquad\qquad\quad 
\stackunder[0pt]{
\includegraphics[width=0.1\textwidth]{ann/ModelB.png}}{}
\quad\qquad\qquad\qquad\qquad\quad 
\stackunder[0pt]{
\includegraphics[width=0.1\textwidth]{ann/ModelC.png}}{}

\stackunder[0pt]{
\raisebox{.4\height}{\includegraphics[width=0.03\textwidth]{ann/Train2.png}}}
\enskip
\subfloat[text][Total accuracy = 98.8125\%]{
\includegraphics[width=0.29\textwidth]{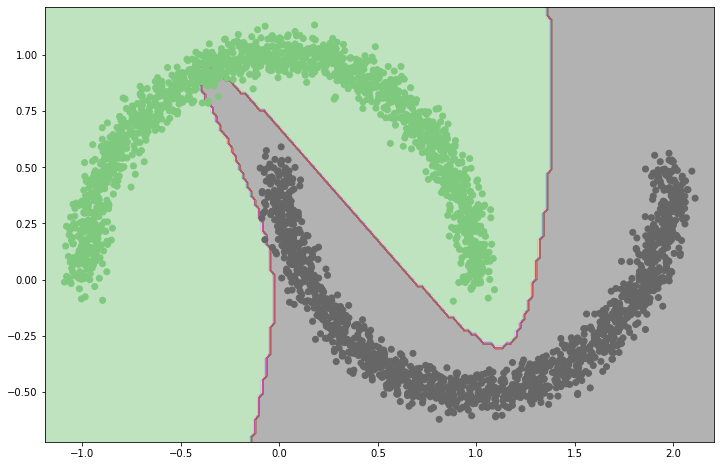}
\label{fig:full_U_seed0_dec_1}}
\quad
\subfloat[text][Total accuracy = 89.8125\%]{
\includegraphics[width=0.29\textwidth]{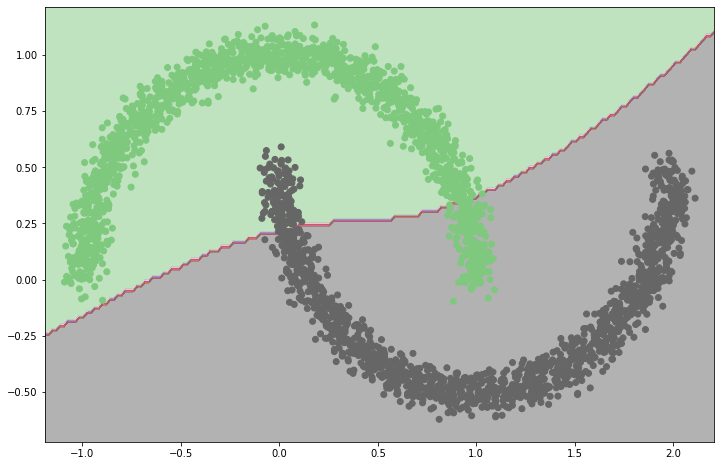}
\label{fig:4_wires_seed0_dec_1}}
\quad
\subfloat[text][Total accuracy = 89.375\%]{
\includegraphics[width=0.29\textwidth]{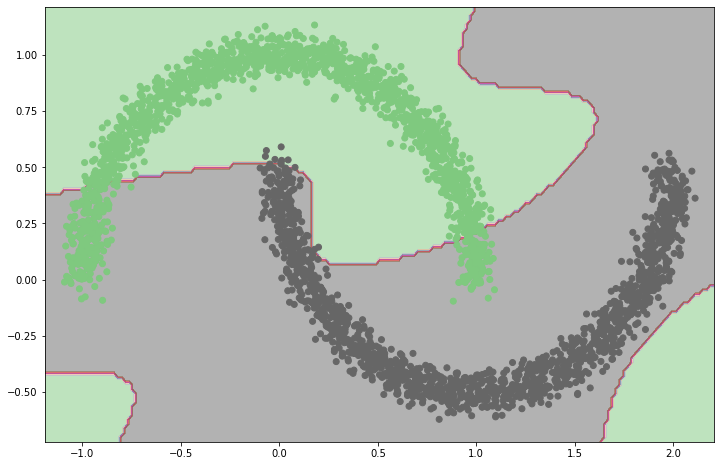}
\label{fig:full_U_seed0_learn_dec_1}}

\stackunder[0pt]{
\raisebox{1.40\height}{\includegraphics[width=0.035\textwidth]{ann/Test.png}}}
\enskip
\subfloat[text][Total accuracy = 98.75\%]{
\includegraphics[width=0.29\textwidth]{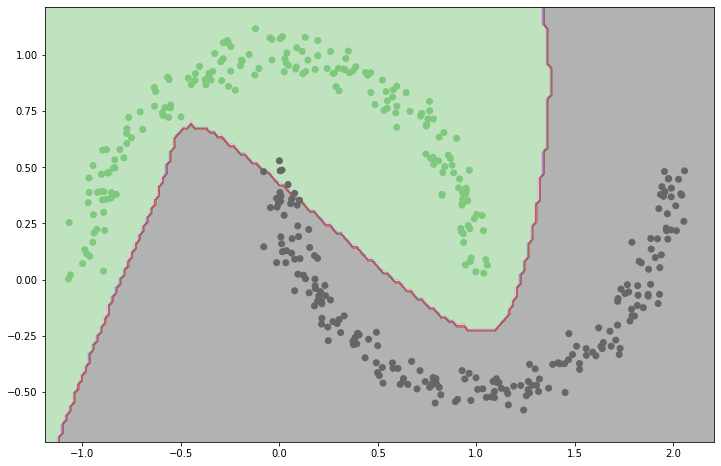}
\label{fig:full_U_seed0_dec_2}}
\quad
\subfloat[text][Total accuracy = 90\%]{
\includegraphics[width=0.29\textwidth]{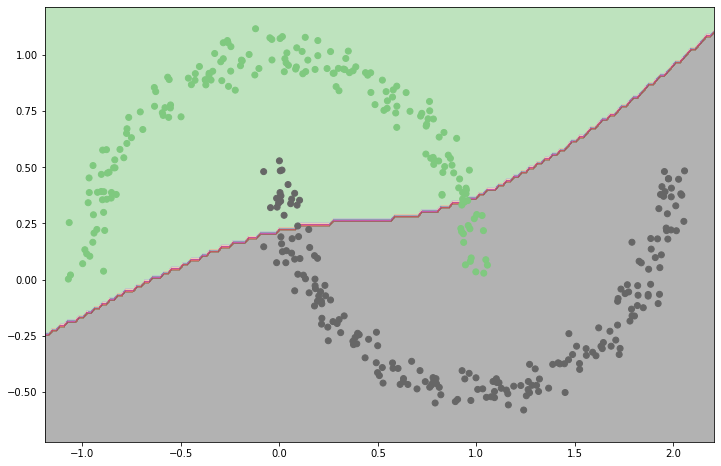}
\label{fig:4_wires_seed0_dec_2}}
\quad
\subfloat[text][Total accuracy = 91.5\%]{
\includegraphics[width=0.29\textwidth]{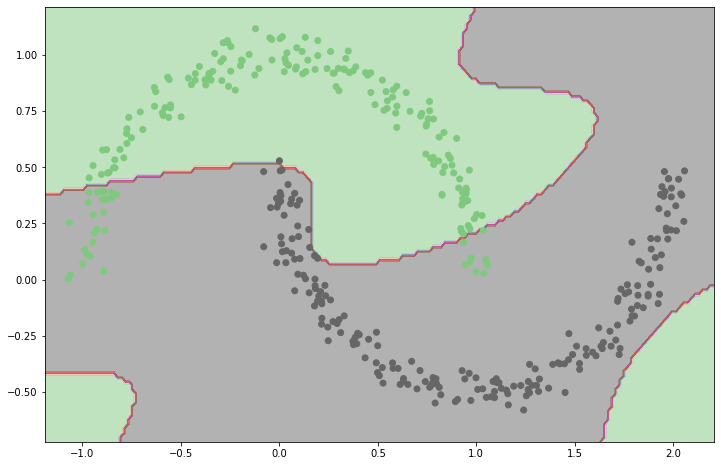}
\label{fig:full_U_seed0_learn_dec_2}}

\stackunder[0pt]{
\raisebox{0.3\height}{\includegraphics[width=0.033\textwidth]{ann/Validation.png}}}
\enskip
\subfloat[text][Total accuracy = 99.25\%]{
\includegraphics[width=0.29\textwidth]{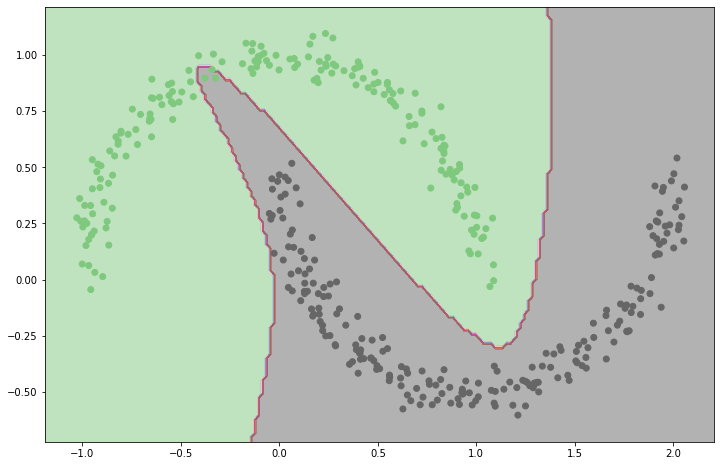}
\label{fig:full_U_seed0_dec_3}}
\quad
\subfloat[text][Total accuracy = 91.25\%]{
\includegraphics[width=0.29\textwidth]{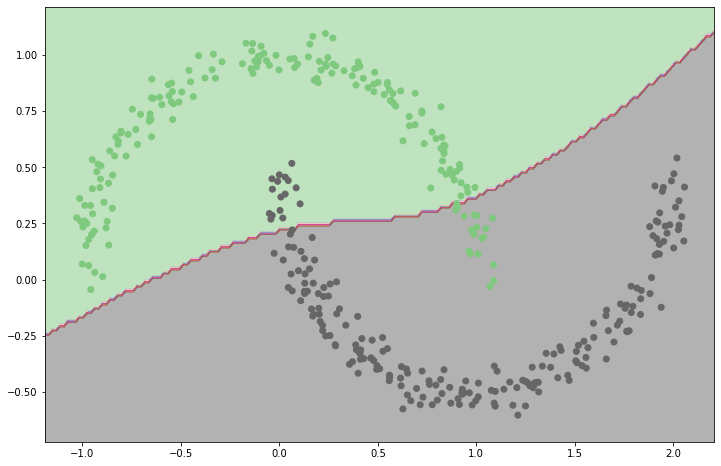}
\label{fig:4_wires_seed0_dec_3}}
\quad
\subfloat[text][Total accuracy = 91\%]{
\includegraphics[width=0.29\textwidth]{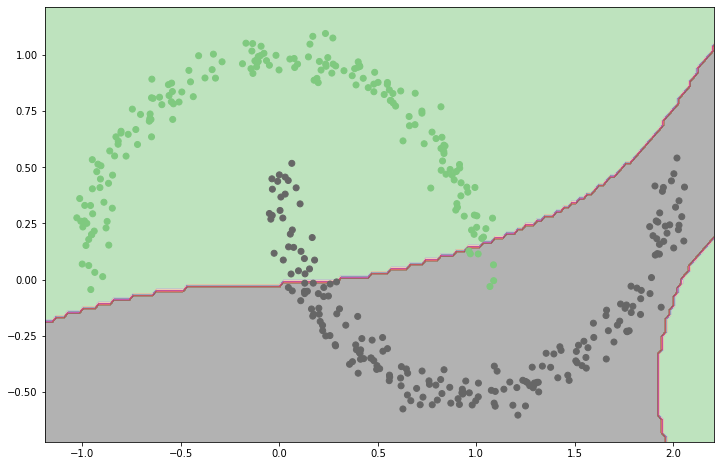}
\label{fig:full_U_seed0_learn_dec_3}}

\caption{
Plots of the contour decision boundary lines given by the gates from QSD after learning the quantum matrices which generated Figure \ref{fig:full_U_seed0}. These plots are expected to look identical to those in Figure \ref{fig:full_U_seed0}, demonstrating that QSD decomposed the unitary matrices (or matrix) correctly. 
}
\label{fig:full_U_seed0_dec}
\end{figure}

QSD decomposes the learned matrix (or matrices) into a set of universal gates whose parameters, gates, and gate order reflect the behavior of the learned unitary matrix (or matrices) on the proper wires. These decomposed gates allow for the creation of the resulting contour plots which produce the decision boundary with the expected accuracy values (as seen in \autoref{fig:full_U_seed0_dec}). QSD completes the pipeline and this algorithm works as expected.

The process of learning the full unitary matrix and then decomposing the learned matrix into unitary gates can be taken a step further by recombining certain sequences of gates into more complex known quantum logic gates (e.g. SWAP, Toffoli, Fredkin, etc). While this recombination step is not shown in this study, such an extra step may reduce the number of implemented gates in the circuit.

\section{Conclusion}
While some fine tuning of the hyperparameters is needed, the proposed and demonstrated algorithm is much more preferable than constructing circuits by hand with ad hoc logic, as such logic assumes a priori knowledge of the underlying features which the user may or may not possess. The algorithm presented in this paper simultaneously discovers a quantum system's architecture and learns the system's parameters. While hyperparameters must be considered, this methodology outlines a path forward when the immediate architecture and parameters of such a system are unknown and a gated quantum machine learning approach is desired. Such quantum methods may improve with the application of more efficient and robust methods of calculating the quantum gradient. The proposed approach makes efficient use of the number of available wires as it maximizes the number of possible learned parameters in the trained unitary matrix.



\section{Acknowledgements}
The author would like to thank Eliana Stefani and Joe Stefani for their insightful conversations. The author would like to thank Walter Vinci and Daniel Davies for their constructive feedback. The author would like to thank Maria Schuld for generously helping the author understand the Pennlyane API backend. The author would also like to thank Raban Iten and Roger Colbeck for their plentiful assistance and guidance on how to use their package, UniversalQCompiler.

\bibliographystyle{unsrt}  

\bibliography{bibtex}


\end{document}